\documentclass[russian,twocolumn,showpacs,preprintnumbers,amsmath,amssymb,showkeys]{revtex4-2}
\usepackage{graphicx}
\usepackage{dcolumn}
\usepackage{bm}
\usepackage{bigints}
\usepackage{babel,amssymb,amsmath,graphicx,hyperref}

\begin{document}

\title{Observing of background electromagnetic radiation of the real sky through 
               the throat of a wormhole}

\author{M.A. Bugaev}
\affiliation{Moscow Institute of Physics and Technology, 9, Institutsky lane, Dolgoprudnyi, Moscow region, 141700, Russia}
\author{I.D. Novikov}
\affiliation{Astro-Space Center of P.N. Lebedev Physical Institute, 84/32, Profsoyusnaya, Moscow, Russia 117997,}
\affiliation{The Niels Bohr International Academy, The Niels Bohr Institute, Blegdamsvej 17, DK-2100, Copenhagen, Denmark}
\affiliation{National Research Center Kurchatov Institute, 1, Akademika Kurchatova pl., Moscow,  Russia 123182}
\author{S.V. Repin}
\affiliation{Astro-Space Center of P.N. Lebedev Physical Institute, 84/32 Profsoyusnaya, Moscow, Russia 117997}
\author{P.S. Samorodskaya}
\affiliation{Moscow Institute of Physics and Technology, 9, Institutsky lane, Dolgoprudnyi, Moscow region, 141700, Russia}
\author{I.D. Novikov, jr.}
\affiliation{Sternberg Astronomical Institute, Moscow State University,13 Universitetsky avenue, Moscow, 119234, Russia}

\begin{abstract}
       The numerical investigation conducted in this paper addresses the problem of CMB radiation imaging as seen
through the throat of the Ellis-Bronnikov-Morris-Thorne wormhole. It is assumed that both throats of the wormhole are 
relatively close to our stellar neighborhood, so close that the view of the ambient background radiation by an observer at 
the other throat of the wormhole is virtually identical to that seen from the Solar System neighborhood. A map of 
the temperature distribution of the cosmic microwave background radiation observed through the mouth of the wormhole 
has been constructed as well as a view of the Milky Way through the mouth of the wormhole. The resultant image
contains characteristic details that enable it to be distinguished from an image produced by a black hole.
\end{abstract}


\maketitle

\section{Introduction}

       Wormholes are hypothetical objects widely discussed in modern day astrophysics and relativistic physical theory. 
As featured by the simplest models, there are two openings (holes) in three-dimensional space, arbitrarily far apart, connected by a 
throat that lies outside our space-time.  In these simplest theoretical models, a wormhole can be considereded as a 
static throat connecting two three-dimensional, asymptotically flat spaces, which we will conventionally denote as space-1 
and space-2.

       The possibility of the existence of wormholes has long been predicted theoretically both within the framework of 
general relativity (see, for example, \cite{Ellis_1973, Bronnikov_1973, Morris_1988a, Morris_1988b, 2015AmJPh..83..486J,
Visser_1989, 2017PhRvD..95b4030T, 2018PhRvD..97h4051T, Javed_2022,2024ChPhC..48f5105H, 2023EPJC...83..284T,
2024ChPhC..48b5104A, 2024GReGr..56...58A,2024PhRvD.109l4065X}) and in alternative theories of 
gravity~(\cite{Agnese_1995, Vacaru_2002, Furey_2005, Eiroa_2008, Botta_2010, DeBenedictis_2012, 
2015PhRvD..91l4020O, Zangeneh_2015, Elizalde_2018, Shaikh_2018, Godani_2020, Singh_2020, Mustafa_2021, Sokoliuk_2022,
Godani_2023}). However, these objects have not yet been discovered in astrophysical observations. 
The possibility of conducting such observations is actively discussed~\cite{Kardashev_2014}. In the 
paper~\cite{2024GReGr..56...52K} it is proposed to extend mathematical methods of constructing black hole shadows 
to wormholes with the metric tensor component $g_{00} = 1$.

       The entrances to static wormholes are similar to the entrances to black holes. Therefore, to set up such observations, 
it is necessary to find specific characteristics that manifest in the observational deviations between such objects. The most 
important distinguishing feature of wormholes is the fundamental possibility of seeing light rays passing through the throat 
from one space to another (say, from space-1 to space-2, or vice versa, from space-2 to space-1). This determines the 
fundamental difference between the observed silhouettes of wormhole entrances and the silhouettes of black holes.
Unlike the black hole silhouettes, inside the wormhole silhouettes one can, in principle, see the images of objects 
located in another space and background electromagnetic radiation from another space. Of course, these images are 
subject to significant changes due to the curvature of the trajectories of the light rays passing through the throat. In the 
papers~\cite{Bugaev_2021, Repin_2022, 2023PhRvD.108l4059B}, the fundamental issues of constructing an image of 
the glow of homogeneous background emission observed from another space are considered. In this paper, in development 
of the issues stated in~\cite{Repin_2022, 2023PhRvD.108l4059B}, we consider the appearance of the real sky image
observed through the throat of a wormhole.

\section{Physical characteristics of a wormhole and computation of light ray trajectories}

        In this paper, we consider the Ellis-Bronnikov-Morris-Thorne static wormhole model, which metric can be 
written as:
\begin{equation}
     ds^2 = dt^2 - \cfrac{r^2}{r^2 - q^2}\,\, dr^2 - r^2
            \left(
               d\vartheta^2 + \sin^2\vartheta \, d\varphi^2
            \right)\,,
          \label{MT_metric2}
\end{equation}
where $q$ is a constant equal to the radius of the wormhole throat. In the equation (\ref{MT_metric2}) the speed of light is 
set to $c=1$, and the radial coordinate~$r$ is chosen so that the circumference is equal to~$2\pi r$. If such radial 
coordinate~$r$ is used to construct the trajectories of light rays, then the trajectories of quanta look more natural and clear.

       It is assumed that the observer is in space-1 and observes the background radiation coming through the wormhole 
from space-2. The sky that is observed through the wormhole is formed by the rays coming to the observer from space-2 
through the throat of the wormhole.  The observer is located in space-1 at a distance from the entrance to the wormhole 
that is much greater than the size of the throat. In addition to these rays, the observer will see the surrounding background radiation in 
space-1, formed by the rays that did not pass through the wormhole. Of course, the rays passing near this entrance will be 
curved, distorting the appearance of the background radiation. We also consider the appearance of the sky (background 
radiation), distorted by this process. These rays form the outer silhouette of the wormhole. The appearance of this outer 
silhouette is also considered.

        We will first consider the view of the sky through the wormhole, and then also present the view of the radiation in 
space-1, in the vicinity of the wormhole's entrance.

        The paper \cite{Kardashev_2020} lists the features by which one can distinguish the image of a black hole from 
the image of a wormhole. In particular, one of the points in there states that the light rays can pass through a wormhole, 
and therefore the observer can see the objects of another asymptotically flat space. The angular size of such images should 
be less than the angular size of the wormhole's throat. Such details can never be observed in the image of a black hole. One 
such image could be a map of the distribution of temperature fluctuations in the cosmic microwave background radiation or 
a map of our Galaxy (the Milky Way) when observed from the vicinity of the Earth.

\begin{figure}[!htb]
  \centerline{
  \includegraphics[width=0.92\columnwidth]{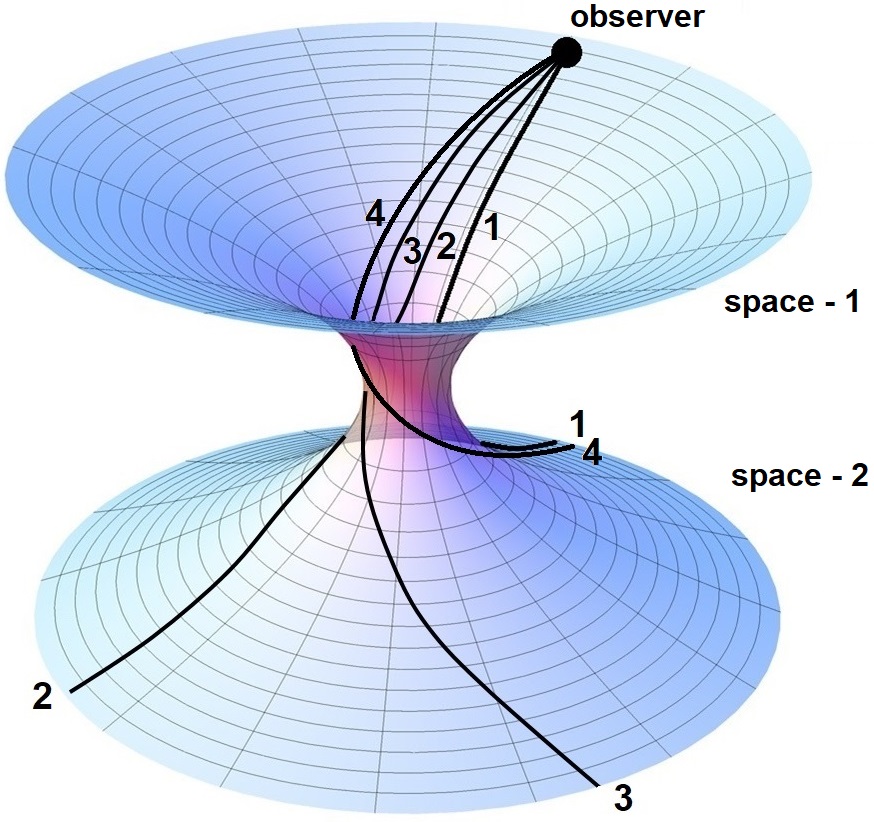}
             }
  \caption{Cross-section of a wormhole for $\theta = \text{const}$. The cross-section shows two asymptotically 
                 flat spaces connected by a wormhole. The observer is located in space-1, and the luminous sky is located 
                 in space-2. The figure schematically shows four photon trajectories with different impact parameters 
                 less than the throat radius, passing through the throat of a wormhole from one space to another.}
  \label{WH_image_1}
\end{figure}

        The location of the observer and the wormhole is shown schematically in Fig.~\ref{WH_image_1} as a
two-dimensional spatial section. We will conventionally assume that the observer is located in space-1, and the radiation 
comes from the celestial sphere, which is in space-2, through the throat of the wormhole.  In this case, the observer is in 
space-1 far from the throat of the wormhole (at "infinity"{}). In addition, Fig.~\ref{WH_image_1} qualitatively shows the 
trajectories of four photons that come to the observer from space-2. Passing through the throat of the wormhole, 
the trajectories of the quanta are curved. This curvature is stronger, the closer the impact parameter of the beam is to 
the radius~$q$ of the wormhole's throat.

        Further, in Section \ref{Temperature_fluctuations}, we will discuss in more detail the relative positions of 
the trajectories and the degree of their curvature.

       To construct the image, it is convenient to assume that light propagates not from the object (the celestial sphere) to 
the observer, but, conversely, from the observer to the object, passing through the mouth of the wormhole. This method 
of computation is justified, since the principle of the ray reversibility is fulfilled in the static metric (\ref{MT_metric2}).

       The equations of photon motion in the field of a wormhole with a metric (\ref{MT_metric2}) are conveniently written in
dimensionless variables: $r' = r/q$~is the radial coordinate, $t' = t/q$~is the time, $L' = L/q$~is the angular momentum 
of the photon, $E$~is its energy at infinity, $Q' = Q/q^2$~is the Carter separation constant (\cite{Carter_1968}), 
$\eta = Q'/E^2$ and $\xi = L'/E$~are the Chandrasekhar constants. In these variables, the equations of motion can be 
written as a system of six ordinary differential equations \cite{Zakharov_1994, Zakh_Rep_1999, Repin_2022}, where 
the primes are omitted for brevity:

\begin{eqnarray}
      \cfrac{dt}{d\sigma} & = & \cfrac{1}{R^2}\,\,, \label{Eq_motion2_1}  \\
      \cfrac{dR}{d\sigma} & = & R_1\,, \label{Eq_motion2_2}  \\
      \cfrac{dR_1}{d\sigma} & = & 2 \left(\eta - \xi^2\right) R^3 - \left(1 + \eta + \xi^2\right) R\,, 
                                \label{Eq_motion2_3} \\
     \cfrac{d\theta}{d\sigma} & = & \theta_1\,, \label{Eq_motion2_4} \\ 
     \cfrac{d\theta_1}{d\sigma} & = & \cfrac{\xi^2\cos\theta}{\sin^3\theta}\,\,,\label{Eq_motion2_5} \\
      \cfrac{d\varphi}{d\sigma} & = & \cfrac{\xi}{\sin^2\theta}\,\,,  \label{Eq_motion2_6}
\end{eqnarray}
where $R=1/r$, $\sigma$~is an independent variable (affine parameter), and $r_1$ and $\theta_1$~are the auxiliary 
variables.

      Methods for numerical solution of the system (\ref{Eq_motion2_1})--(\ref{Eq_motion2_6}) are discussed in 
the papers \cite{Bugaev_2021,Repin_2022,2023PhRvD.108l4059B,Bugaev_2022c}.

\section{Temperature fluctuations through the throat of a wormhole}

\label{Temperature_fluctuations}

       Let space-2 contain the cosmological microwave background distribution of temperature fluctuations. Our first task is 
to construct an image of the sky of space-2 as seen through the throat of the wormhole by an observer located in space-1.

       Let us first consider the general ideas of image formation. As we already noted above, it is more convenient 
to consider the trajectories of quanta that propagate from the observer to the source, i.e. from the observer to various 
points of the celestial sphere in \mbox{space-2}. 

        Fig.~\ref{WH_image_1} shows schematically the cross-section of the wormhole at $\theta = \text{const}$ and 
the trajectories of several quanta with different values of the impact parameter. Trajectory 1 corresponds to a quantum with 
zero impact parameter, i.e. a quantum directed exactly to the center of the wormhole throat. Trajectory 2 corresponds to 
a quantum with a larger impact parameter, at which it turns within the throat of the wormhole by $180^\circ$. The numerical 
solution of the system~(\ref{Eq_motion2_1})--(\ref{Eq_motion2_6}) shows that this occurs at the impact parameter value 
$b = 0.793q$, where $q$~is the throat radius. Trajectory 3 deviates even more, which eventually turns in the throat by about
$270^\circ$. And again the numerical solution of the system~(\ref{Eq_motion2_1})--(\ref{Eq_motion2_6}) allows us to find out 
that this occurs at the impact parameter value of about $b = 0.93q$.  Finally, trajectory 4 makes a full turn in the mouth of 
the wormhole and turns $360^\circ$. The impact parameter in this case has a value $b=0.986q$. This last trajectory is most
interesting because, by reversing the path of the rays, we obtain an interesting conclusion that rays 1 and 4 come out from 
the same point on the celestial sphere.  In other words, the same object when observed through the mouth of a wormhole can 
have several images. It is easy to understand that all possible rays of type 2, i.e. with the same impact parameter, each having 
made half a turn, will also hit the same point on the celestial sphere.

\begin{figure}[!htb]
  \centerline{
  \includegraphics[width=\columnwidth]{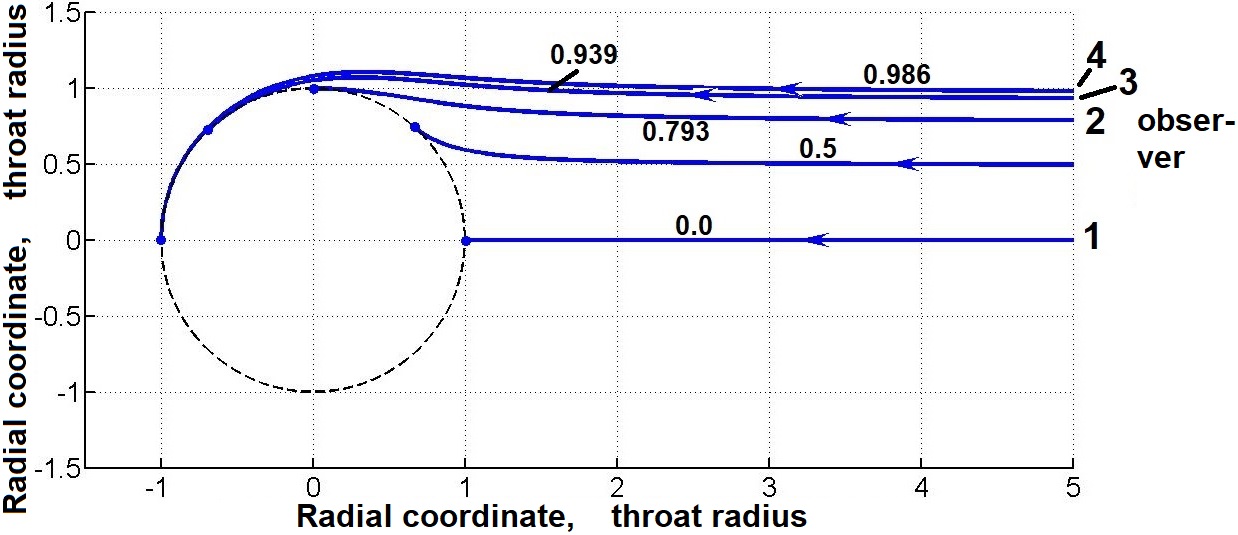}
             }
  \caption{The trajectories indicated in the Figure by numbers 1, 2, 3 and 4 are the same trajectories as in 
                 Fig.~\ref{WH_image_1}, but constructed in space-1 using the numerical integration. Each trajectory 
                 has its number and impact parameter, less than one. The radius of the dotted circle corresponds to 
                 the radius of the wormhole throat. The trajectory with the impact parameter 0.5 is added for clarity. 
                 It lies between the trajectories 1 and 2.}
  \label{WH_image_2}
\end{figure}

\begin{figure}[!htb]
  \centerline{
  \includegraphics[width=\columnwidth]{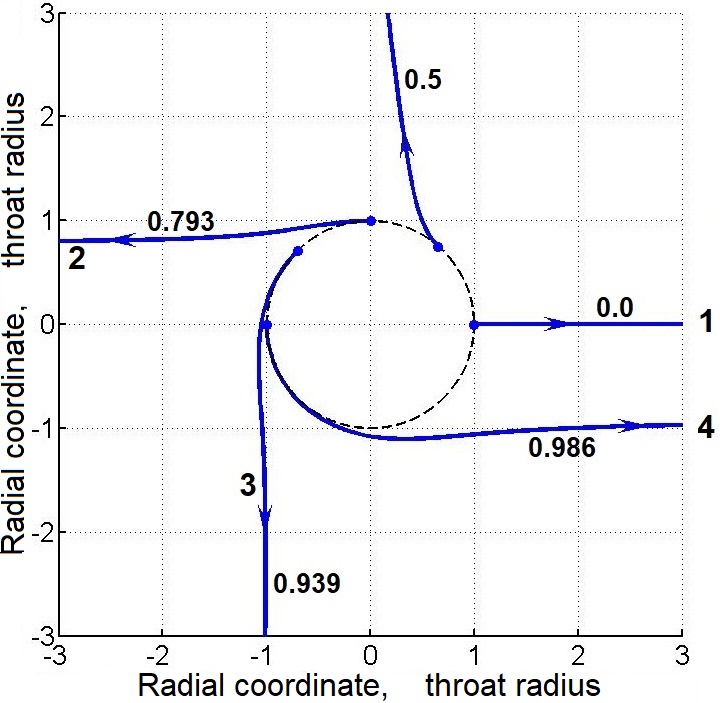}
             }
  \caption{Continuation of the trajectories of Fig.~\ref{WH_image_2} in space-2 after passing the throat. 
                 The trajectories indicated in the Figure by numbers 1, 2, 3 and 4 are the same trajectories as in 
                 Fig.~\ref{WH_image_1},\ref{WH_image_2}, but constructed in space-2 using the numerical integration.
'                Each trajectory has its number and impact parameter, less than one. The radius of the dotted circle 
                 corresponds to the radius of the wormhole throat. The trajectory with the impact parameter 0.5 
                 is added for clarity. It lies between the trajectories 1 and 2.}
  \label{WH_image_3}
\end{figure}

        Fig.~\ref{WH_image_2}--\ref{WH_image_3} decicts five trajectories  four of which coincide with `drawn' trajectories
shown in Fig.~\ref{WH_image_1}, however quanta paths in Fig.~\ref{WH_image_2}--\ref{WH_image_3} is a result of 
numerical solusion of Equations of motion (\ref{Eq_motion2_1})--(\ref{Eq_motion2_6}). The fifth trajectory with 
the impact parameter $b = 0.5q$ is shown for clarity. 
In this case, part of the trajectory from the observer to the wormhole lies in space-1, this part is shown in 
Fig.~\ref{WH_image_2}. The rest of the trajectory (after passing the throat) lies in space-2 and is shown in
Fig.~\ref{WH_image_3}, i.e. the trajectories in Fig.~\ref{WH_image_3} are the continuation of the trajectories shown 
in Fig.~\ref{WH_image_2}.  Next to each trajectory, its number and the corresponding impact parameter are indicated. For 
example, trajectory~2 turns at an angle close to~$180^\circ$ in the throat of the wormhole, and trajectory~3~-- at 
an angle close to~$270^\circ$. Trajectory 1 can be conventionally called a "straight"{} trajectory, which "does not deviate"{}
in the throat. It would especially like to emphasize that it is trajectory~1 that goes to the center of the wormhole and is
"straight"{}, while trajectory~2 turns by~$180^\circ$, although visually it does not look that way.

         It is clear that there will be such rays that can make several turns in the throat of the wormhole. All of them will have 
the impact parameter very close to the radius of the throat $q$. And this means that the image of such an object will contain 
ring structures. The details of the formation of such structures and their appearance are discussed in 
paper~\cite{2023PhRvD.108l4059B} for the case of completely homogeneous background radiation. The distribution of 
the brightness of the image along the radial coordinate is also given there.

        The light rays are also bent when passing through the outer side of the wormhole. This means that ring structures 
and highly distorted astrophysical objects will also be visible from the outer side of the silhouette and the location of these
objects is strictly confines to our Universe. The reason for these distortions is completely analogous to that which occurs for rays 
passing through the throat.
         
\begin{figure*}[!htb]
  \centerline{
  \includegraphics[width=15cm]{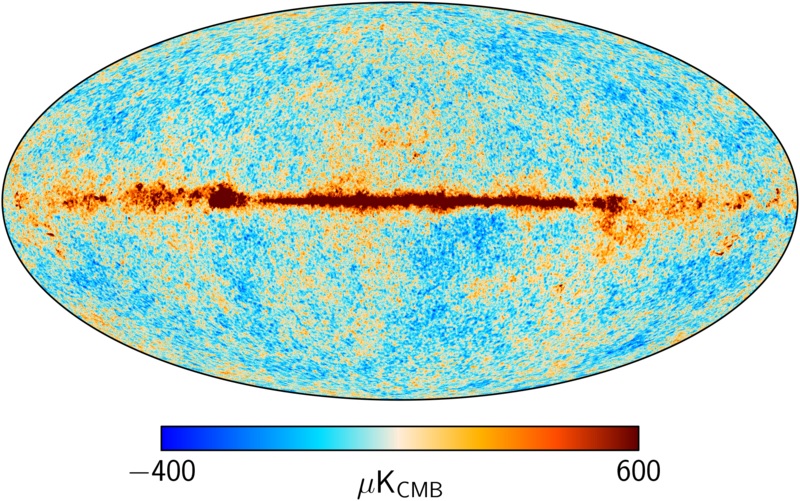}
             }
  \caption{The distribution of microwave radiation temperature over the celestial sphere in galactic coordinates based on 
                 the results of the Planck observatory \cite{2016A&A...594A...9P}.}
  \label{Planck_CMB_map}
\end{figure*}

      From the above qualitative considerations it becomes clear that the images of astrophysical objects when observed through 
the throat of a wormhole will appear highly distorted.

\begin{figure*}[p]
  \centerline{
  \includegraphics[width=12cm]{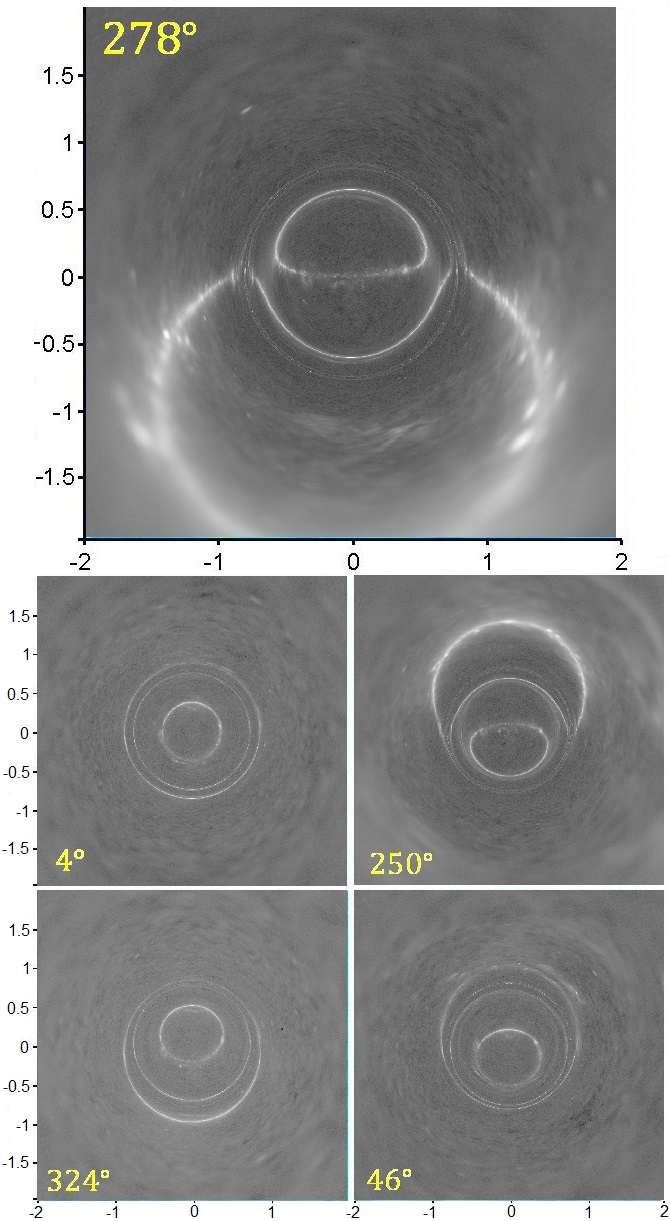}
             }
  \caption{Map of the temperature fluctuations of the cosmic microwave background radiation as seen through the throat 
                 of a massless wormhole and the immediate vicinity of the wormhole. Black and white palette.
                 The radial coordinate is given in units of the wormhole throat radius. The angle between the perpendicular 
                 to the Galactic plane and the observer's line of sight is also given.}
  \label{CMB_bw_Image}
\end{figure*}        

\begin{figure*}[p]
  \centerline{
  \includegraphics[width=12cm]{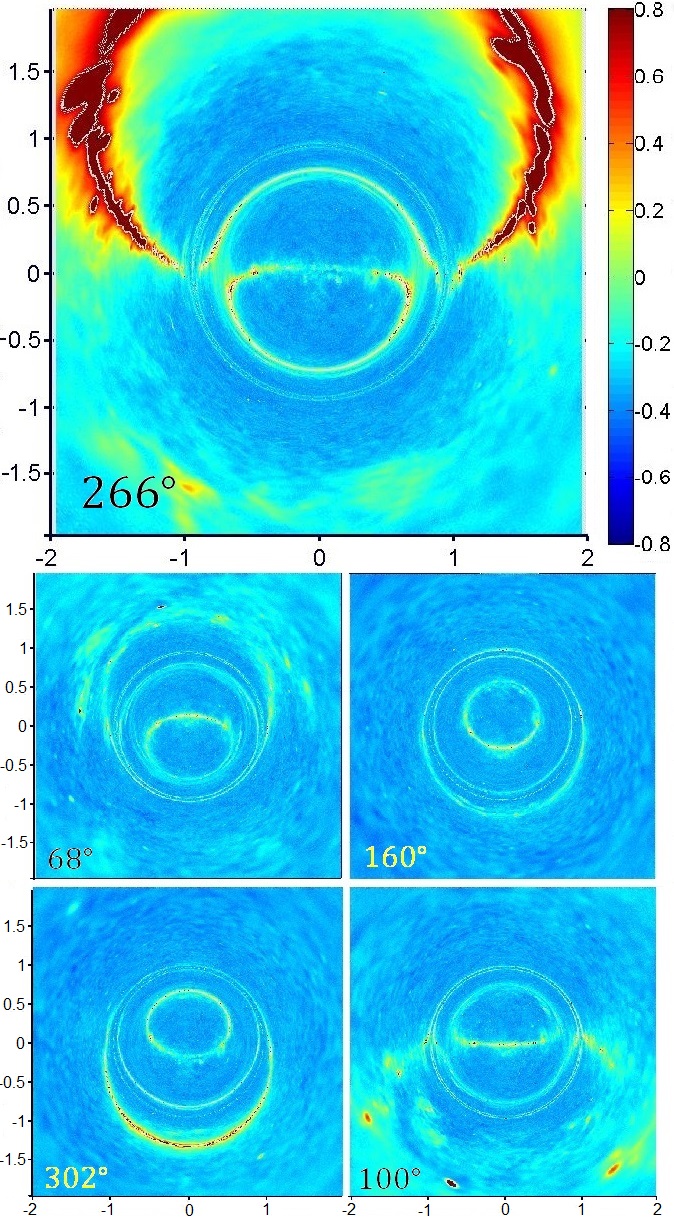}
             }
  \caption{Map of the temperature fluctuations of the cosmic microwave background radiation as seen through the throat 
                of a massless wormhole and the immediate vicinity of the wormhole. Color palette. The radial coordinate is given 
                in units of the wormhole throat radius. The angle between the perpendicular to the galactic plane and 
                the observer's line of sight is also given.}
  \label{CMB_col_Image}
\end{figure*}        

       A typical view of the distribution of temperature fluctuations on the celestial sphere according to Planck data is 
shown in Fig.~\ref{Planck_CMB_map} (see \cite{2016A&A...594A...9P}). A uniform scale of temperature fluctuations was 
used for the construction. This image is widely known to specialists. The bright horizontal strip in the image is our Galaxy. 
The data for constructing such maps can be found in the database of this space mission.

       Fig.~\ref{CMB_bw_Image} shows an image of the temperature fluctuations of the cosmological misrovawe background 
rediation when observed through the throat of the Ellis-Bronnikov-Morris-Thorne wormhole. The image is shown from both 
the outer and inner sides of the wormhole. In this case, we assume that the other exit of the wormhole is in our Universe and 
lies not very far from the entrance~\cite{Kardashev_2020}, so that the view of the sky in our space is almost the same for 
the observers near both entrances. The observer's line of sight angle is measured between the normal and the Galaxy plane, 
lies in the plane passing through the center of the Galaxy and varies from $0^\circ$ to $360^\circ$. The value of this angle 
is indicated on each panel of Fig.~\ref{CMB_bw_Image}. The image in black and white color grading depicts temperature 
fluctuations on a logarithmic scale  Using such a scale, the naked eye better distinguishes the characteristic large details of 
the image. The image is similar to that shown above in Fig.~\ref{Planck_CMB_map}. In Fig.~\ref{CMB_bw_Image}, 
the galactic plane distorted by lensing is clearly visible. Near the boundary of the visible silhouette of the throat, these 
distortions are especially noticeable.

       Fig.~\ref{CMB_col_Image} shows more images of microwave temperature fluctuations, seen both through the wormhole 
throat and in the surrounding space, but rendered in a color (geographical) palette. A logarithmic scale of temperature 
fluctuations is used here too, but zero point of this scale does not refer to any particular value. Note that the different 
palettes carry different information about the object and allow us to extract visually the different characteristic details of 
the image. Large details are clearly visible in a black and white image, while the small structures are easier to identify in 
a color palette.

      One can also estimate the total expected radiation flux in the wormhole observations.  Let us assume that 
a conventional radio telescope observes a source at a frequency of 240~GHz (the maximum flux of the CMB radiation falls at 
a frequency of 160~GHz). If we assume that the angular size of the throat of a wormhole coincides with the angular size of 
the shadow of the black hole in the center of the Galaxy (source Sgr~A$^*$), then the flux from such a source can reach 
100~$\mu$Jy in order of magnitude.
Such an object can be detected by modern radio telescopes. The image should include a bright central part and very bright 
photon rings encircling the central part. However, the width of the photon rings, starting from the second one, turns out to be 
smaller than the pixel size and it is impossible to depict them on such a scale. In the bright central part, one could try 
to detect the temperature fluctuations on very small angular scales. To observe this image, of course, a high angular resolution 
is required. Such a resolution can theoretically be achieved if, for example, one of the antennas is located at the Lagrange 
points $L_1$, $L_3$ of the Earth--Moon system or at the Lagrange points $L_2$, $L_4$, $L_5$ of the Earth--Sun system. 
However, there are still many technical challenges to overcome here \cite{Mikheeva_2020,2022aems.conf..292M}. One can 
hope that these observations can be implemented in future space missions.

\section{The starry sky (Milky Way) observed through the throat of a wormhole}

\begin{figure*}[p]
  \centerline{
  \includegraphics[width=12cm]{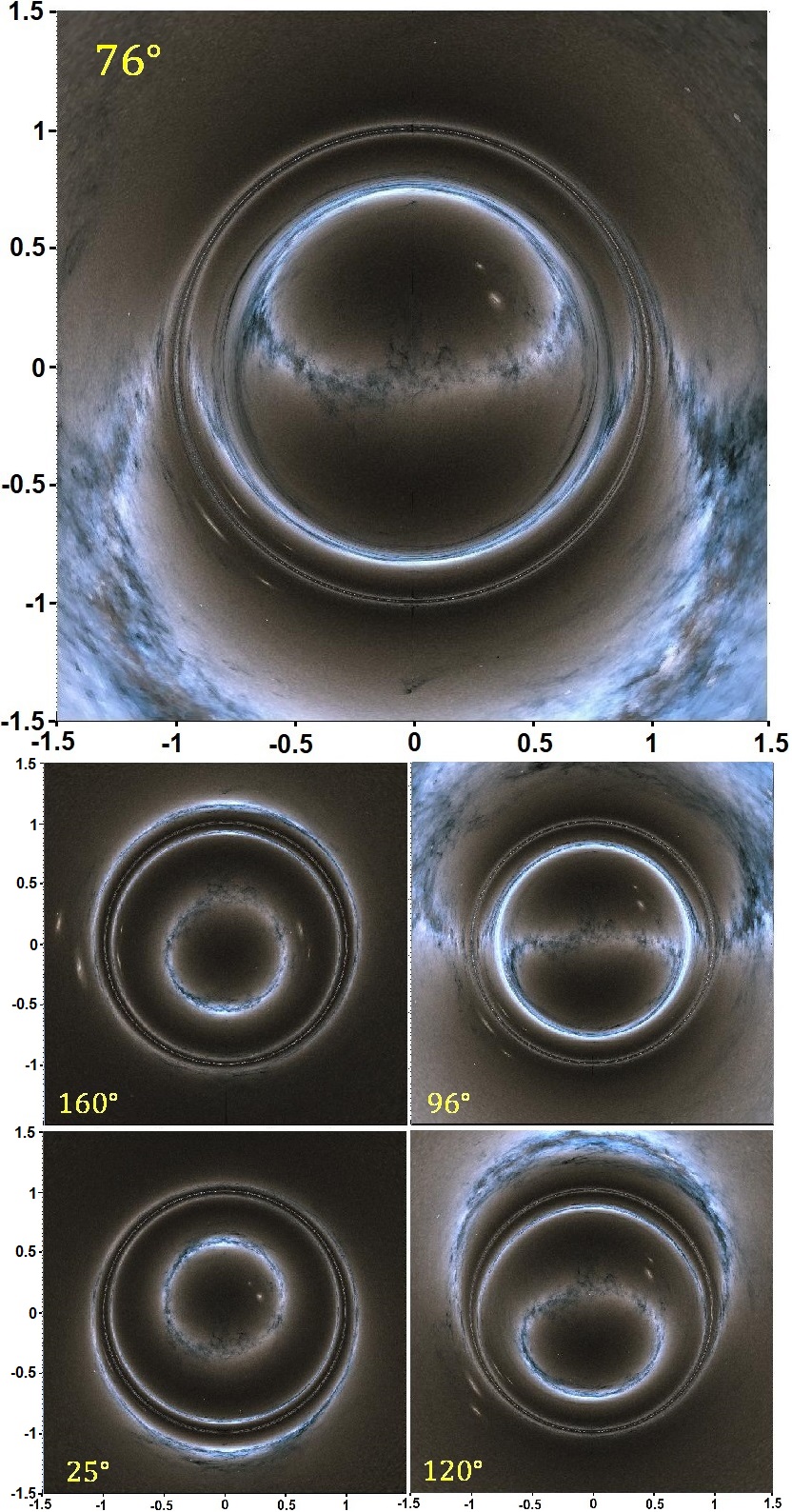}
             }
  \caption{Map of the Milky Way as seen through the throat of a massless wormhole and the immediate 
                 vicinity of the throat. The image is shown in natural colors. The radial coordinate is given in units 
                 of the wormhole throat radius.}
  \label{MWay_bw_Image}
\end{figure*}        

       In addition to the temperature fluctuations of the microwave background, it is also possible to depict a general view 
of the starry sky in the optical range, observed through the throat of the wormhole. If the second entrance of the wormhole 
is in our Universe and even in our Galaxy, relatively close to the observer, then the main visible detail in the sky in 
the optical range will be the Milky Way. The image of the Milky Way, visible both through the mouth of the wormhole and 
from the outside of the mouth, is shown in Fig.~\ref{MWay_bw_Image}.  The angle between the perpendicular to 
the Galactic plane and the observer's line of sight is indicated on each panel and is similar to that used in 
Figs.~\ref{CMB_bw_Image} and~\ref{CMB_col_Image}.

       The appearance of the sky depends, of course, on the position of the entrance. Fig.~\ref{MWay_bw_Image} shows 
the case when the entrance is located near the galactic plane relative to the observer. The Figure clearly shows that 
the distorted image of the Milky Way has an angular size of the same order as the angular size of the throat. In addition, 
the familiar features such as the Magellanic Clouds, the center of the Galaxy, and known features of the Milky Way image 
are clearly visible in the image. Thus, at an angle of~$76^\circ$, the central part of the Galaxy (a bright arc) is clearly visible 
inside the silhouette of the wormhole in the upper part, and both Magellanic Clouds are visible to the right and above 
the center. At an angle of $25^\circ$, the image of the Galaxy is almost an exact circle, shifted relative to the image center, 
but the galactic center still stands out well compared to the opposite part of the Galaxy. You can even distinguish a dark (dust) 
band inside the Milky Way. At an angle of $76^\circ$, $96^\circ$ and $120^\circ$, the lensed parts of the Galaxy on the outer 
side of the wormhole silhouette are clearly visible in the image. At the same time, to the left and below the center, you can 
again distinguish another image of the Magellanic Clouds. 

     Such details could be seen when observing a wormhole if it is located near the Solar System. And this, in turn, means that 
the detailed structure of this image can be obtained and identified by modern interferometers in the optical range at the limits 
of their resolution \cite{Mikheeva_2020, 2022aems.conf..292M,2023AJ....166..123K, 2023AJ....165...41B,2024AmJPh..92...43R}. 
It should be especially emphasized that near the visible silhouette of the wormhole's throat there are characteristic ring structures, 
which are not just photon rings, but nothing more than a distorted image of the entire Milky Way. Such structures are 
fundamentally absent in the shadows of black holes \cite{2024ARep...68....1M} and, therefore, the presence of these 
structures in the image allows us to conclude that the observed object may be a wormhole.

      Fig.~\ref{MWay_bw_Image}, as well as Fig.~\ref{CMB_bw_Image}--\ref{CMB_col_Image}, show only the image 
distortions, but do not take into account its brightness. This representation allows us to better identify the relativistic effects 
associated with lensing and image distortion. It turns out that the correct accounting of brightness has little effect on the overall 
appearance of the image. The brightness of the inner part of the image up to 0.7 of the wormhole throat radius can be 
considered as a constant with acceptable accuracy~\cite{2023PhRvD.108l4059B}.

\section{Conclusion}

      Wormholes with close entrances may possibly be present in the centers of galaxies \cite{Kardashev_2020}.

       To observe the wormholes, it is necessary to know the characteristic features of the images by which these objects 
could be distinguished from black holes or any other relativistic objects. Simulating these images allows us to identify such 
features by which an object could be confidently classified as a wormhole.

       In the considered model of the Ellis-Bronnikov-Morris-Thorne wormhole image, there are characteristic details that can 
be used to observe and identify these objects in the interferometric observations~\cite{Mikheeva_2020,2022aems.conf..292M}. 
Thus, inside the silhouette of the wormhole one could observe an image of the Milky Way, heavily distorted by lensing.
In this image, it would be possible to pick out and identify the familiar parts of the image, such as the Magellanic Clouds or 
the "coal sacks". The appearance of this image itself depends, of course, on the position of the observer and on the position 
of the second entrance to the wormhole. Since the central part of the image has approximately the same brightness, then 
to observe such a picture, which is shown in Fig.~\ref{MWay_bw_Image}, the resolution of an interferometer with a base of
$10^6$~--~$10^7$~km and a frequency of 200~--~600~MHz should be sufficient. Then, theoretically, one of the antennas 
of such a device could be located on the Earth, and the other at the Lagrange point $L_2$ of the Sun--Earth system or 
the Earth--Moon system. It would also be possible to use a telescope installed on the Moon. The resulting images after 
processing can be compared with the view of the starry sky from the vicinity of the Earth.

      The discovery of wormholes in observations would certainly be an epochal event in astrophysics.
       
\section{Acknowledgments}

      S.R. expresses gratitude to O.N. Sumenkova, R.E. Beresneva, and O.A. Kosareva for the opportunity to fruitfully work 
on this problem. All authors express gratitude to Dr. S.V.Pilipenko for assistance in working with the data and discussing 
the results.

\newcommand{\aaa}{Astron. and Astrophys. }
\newcommand{\aap}{Astron. and Astrophys. }
\newcommand{\aj}{Astronomical Journal }

\bigskip

\bibliography{WH_CMB_AR3}

\end{document}